\documentclass[11pt,twoside]{article}

\usepackage{asp2006}
\usepackage{graphicx}

\begin{document}
\title{Intrinsic Properties of Low-z SDSS Galaxies}   
\author{Ariyeh H. Maller$^1$, Andreas A. Berlind$^{2,3}$, Michael Blanton$^3$, and\\ David W. Hogg$^3$}   
\affil{$^1$Physics Dept., New York City College of Technology, New York, NY\\
	$^2$Physics Dept., Vanderbilt University, Nashville, TN\\
	$^3$CCPP, New York University, New York, NY }    

\begin{abstract} 
Galaxies are seen from different viewing angles and their \emph{observed} properties change as a function of viewing angle.  In many circumstances we would rather know the \emph{intrinsic} properties of galaxies --- those properties
that do not depend on viewing angle.  For a large sample of galaxies it is 
possible to recover the intrinsic properties of galaxies, statistically, by looking for
correlations with galaxy inclination, and then applying a correction to remove
those dependencies. Studying the intrinsic properties of galaxies can give a different impression of the galaxy population and help avoid the mistake of connecting observed properties to quantities that don't depend on inclination like halo mass.
\end{abstract}

\vspace{-0.5cm}
\section*{Inclination Corrections for Low-z SDSS Galaxies}
In order to understand the formation and evolution of galaxies it is necessary to measure the distribution of galaxy properties and how they evolve with redshift.  However, most studies of galaxies look at the distributions of galaxies' 
observed properties like luminosity, color, size and morphology.  Because of
attenuation by dust these properties can change with galaxy inclination and 
therefore what is measured is a convolution of a galaxy's intrinsic properties 
and the effects of dust.  This complicates attempts to understand how galaxy properties evolve with redshift as any observed change  could be due to variations in dust properties or evolution of galaxies' intrinsic properties.  Furthermore, comparison to theory is made more difficult because a model 
of attenuation from dust as a function of inclination is needed to compare 
theoretical models to observations.  

For these reasons it is useful to determine inclination corrections for galaxies.  This can be done for large samples in a statistical manner by searching for
correlations between a galaxy property and inclination, and then assuming that
correlation should be removed \citep{giov:94,giov:95,tully:98,mgh:03,shao:07}.
In \citet{mbbh:08} we perform such an analysis on a sample of galaxies from 
the NYU-VAGC \citep{blan:03c} which are imaged in both the Sloan Digital Sky Survey \citep{york:00} and the Two Micron All Sky Survey \citep{skru:06}. This
gives us $10,340$ galaxies with eight wavebands of coverage from $u$ to $K_s$ with which to determine inclination corrections.

\begin{figure}[t]
   \includegraphics[width=\textwidth] {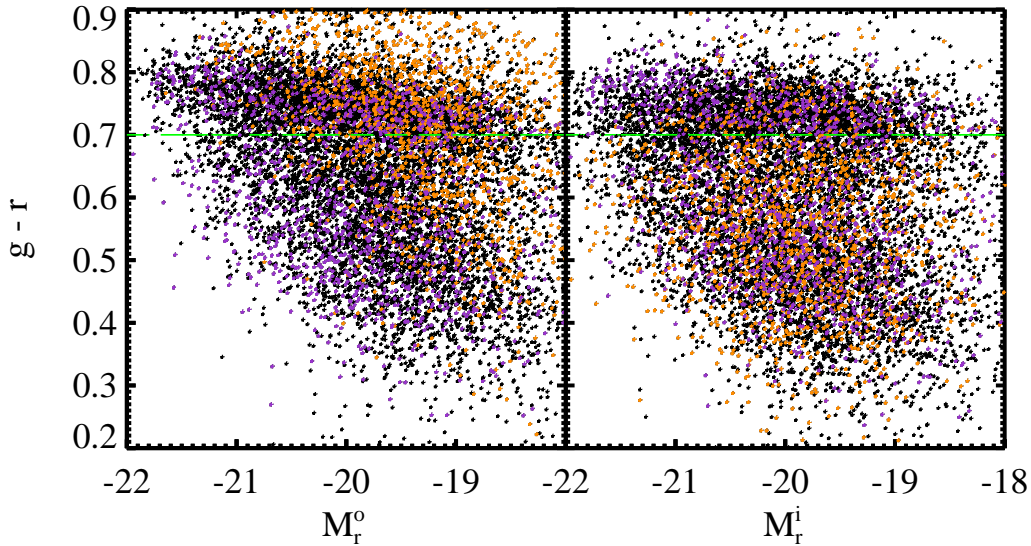} 
   \vspace{-1.6cm}
   \caption{Color-magnitude diagrams for the observed (left panel) and intrinsic
    (right panel) galaxies in our sample. Purple points are face-on ($b/a > 0.85$)
     while orange points highly inclined ($b/a < 0.35$).  Clearly, the observed 
     properties of face-on and edge-on galaxies differ, with edge-on galaxies
     being fainter and redder. However, when inclination corrections are applied
     the two show a comparable distribution in the color-magnitude diagram.  Also,
     the fraction of galaxies that one would consider red decreases when going
     from observed to intrinsic color.}
   \label{fig:cmd}
\end{figure}
When inclination corrections are applied to the sample interesting changes to the population of galaxies can be noted.  Figure \ref{fig:cmd} shows the color-magnitude diagram for observed (left panel) and intrinsic (right panel) galaxy properties.  
The color-magnitude diagram changes in a number of ways.  Galaxies
redder than the red sequence are found in the observed color-magnitude 
diagram, but not once inclination corrections have been applied.  Also, the
fraction of galaxies with $g - r \ge 0.7$ in the observed color-magnitude diagram
is $46\%$.  This drops to $32\%$ when going to the intrinsic color-magnitude diagram a reduction of almost one third in the number of red galaxies a significant change. Please see \citet{mbbh:08} for more detail.

\acknowledgements AHM acknowledges support from the PDAC and the College of Arts and Sciences to attend this meeting.

\end{document}